\documentclass[dvips]{article}

\usepackage{icrc2011}

\title{The VERITAS Extragalactic non-Blazar Program}
\shorttitle{Galante, VERITAS Extragalactic Program}

\authors{Nicola Galante$^{1}$, for the VERITAS Collaboration$^{2}$}
\afiliations{$^1$Harvard-Smithsonian Center for Astrophysics\\ 
$^2$http://veritas.sao.arizona.edu }
\email{ngalante@cfa.harvard.edu}

\abstract{VERITAS is an array of four 12-m diameter imaging atmospheric-Cherenkov 
telescopes located in southern Arizona. Its aim is to study the very high energy 
(VHE: $E>100$~GeV) $\gamma$-ray emission from astrophysical objects. 
In addition to the study of blazars, the VERITAS extragalactic science program
develops a comprehensive observational program of extragalactic non-blazar sources.  
The study of Active Galactic Nuclei (AGN) is intensively pursued through the large MWL 
observational campaign on non-blazar radio galaxies. The success of these MWL 
campaigns has for the first time provided insights to the inner structures of jets 
responsible for $\gamma$-ray emission. 
The problem of the origin and acceleration of ultra-high energy cosmic rays
is pursued through the observation of the starburst galaxy M 82, 
whose detection at VHE delivers important insights on the possible acceleration mechanisms.
Finally, investigation of globular clusters places important 
limits on the millisecond pulsar contribution to their addressed $\gamma$-ray emission. The 
VERITAS extragalactic non-blazar science program and its results are presented.}

\begin{document}
\maketitle

%Begin the section.
\section{Introduction}

The largest population of VHE-detected $\gamma$-ray sources
are blazars, a sub-category of AGNs in which the ultra-relativistic jet, produced by the
accretion of matter around a super-massive black hole, is aligned within a few degrees to the
observer's line of sight~\cite{UrryPadovani}. The most commonly accepted model
to account for the $\gamma$-ray emission from the jet of AGNs is inverse-Compton
scattering of the synchrotron photons produced by shock-accelerated electrons and positrons
within the jet itself~\cite{Jones1974}. In these aligned sources relativistic beaming substantially
boosts the apparent flux.
%Hadronic interactions producing neutral pions which decay into photons~\cite{Mannheim1992}, 
%and synchrotron emission from protons~\cite{Aharonian2000} are also possible scenarios for the high-energy 
%component of blazars. 
Non-blazar AGNs are typically oriented at 
larger angles from the observer's line of sight, becoming much more challenging. However,
the misalignment enables imaging of the jet's structure, crucial to identifying the emission regions
and probing models of the acceleration mechanism. Given the typical angular resolution
of the order of several arcminutes in $\gamma$-ray instruments, jet substructures are not resolved
in the $\gamma$-ray energy band, but  they are in other wavelengths.
Correlation studies through coordinated
multi-wavelength (MWL) observational campaigns on radio galaxies are a viable strategy
to investigate the physical processes at work in the substructures of the jet.

The advantage of observing radio galaxies is that it is possible to study also the rich environment
in which they are typically located. It has been seen that radio galaxies are preferentially
located in cluster of galaxies~\cite{Prestage1988}. Their powerful jets energize the intra-cluster medium through the 
termination shocks accompanied by particle acceleration and magnetic field amplification. 
Large scale AGN jets and cluster of galaxies are believed to be potential accelerator for cosmic 
rays~\cite{Dermer2009}, therefore the modeling of the dynamics of both populations is of particular 
interest for the cosmic-ray community.

Beside AGN-related environments,
starburst galaxies are also good candidates as ultra-high energy cosmic rays accelerators. The active
regions of starburst galaxies have a star formation rate about 10 times larger than the rate
in normal galaxies of similar mass, with a consequent higher rate of novae and supernovae.
The cosmic rays produced in the formation, life, and death of their massive stars are
expected to eventually produce diffuse gamma-ray emission via their interactions with interstellar 
gas and radiation.

Finally, globular clusters are the closest extragalactic structures whose physics is interesting
to the $\gamma$-ray community. They can host hundreds of millisecond pulsars
which can accelerate leptons at the shock waves originating in collisions of the pulsar winds 
and/or inside the pulsar magnetospheres. Energetic leptons diffuse gradually through the 
globular cluster. Comptonization of stellar and microwave background radiation is therefore
expected to be responsible of $\gamma$-ray emission.

The indirect search for dark matter (DM) candidates,
 is also part of the VERITAS extragalactic non-blazar program. A dedicated contribution
on the VERITAS DM program is presented in a separate proceeding.
Highlights on the research topics and results of the VERITAS extragalactic non-blazar science
program are here presented.

\section{The VERITAS Instrument}

The VERITAS detector is an array of four 12-m diameter imaging
atmospheric Cherenkov telescopes located in southern Arizona~\cite{Weekes}. 
Designed to detect emission from astrophysical
objects in the energy range from 100~GeV to greater than 30~TeV,
VERITAS has an energy resolution of $\sim$15\% and an angular
resolution (68\% containment) of $\sim$0.1$^\circ$ per event at 1~TeV.  
A source with a flux of 1\% of the Crab Nebula flux is detected in $\sim$25~hours
of observations, while a \mbox{5\% Crab Nebula} flux source is detected in less than
2~hours.  The field of view of the VERITAS telescopes is
3.5$^\circ$.  For more details on the VERITAS instrument and the imaging atmospheric-Cherenkov
technique, see~\cite{Perkins2009}.

\section{The Extragalactic non-Blazar Science Program}

\subsection{Radio Galaxies}

Most of the VERITAS observations of radio galaxies are on the radio galaxy M~87.
This AGN is located in the center of the Virgo cluster at a distance of $\sim$16~Mpc
and is currently the brightest detected VHE radio galaxy.
M~87 was originally detected with marginal significance by HEGRA at TeV 
energies~\cite{Aharonian2003}, 
and later also by HESS~\cite{Aharonian2006}, VERITAS~\cite{Acciari2008} and MAGIC~\cite{Albert2008}.
This giant radio galaxy has always been of particular interest because
its jet lies at $\sim$20$^\circ$ respect to the line of sight and
its core and the structure of the jet
are spatially resolved in X-ray, optical and radio observations,
thus it is an ideal candidate for correlated MWL studies~\cite{Wilson2002}.

In 2008 VERITAS coordinated an observational campaign with two other major VHE observatories
(MAGIC, HESS), overlapping with VLBA radio observations~\cite{M87Science}. 
Three Chandra X-ray pointed observations have also been performed during the first half of 2008.
Multiple flares at VHE have been detected. In X-rays, the inner-most knot in the jet (HST-1) was found in low
state, while the core region was in high state since 2000. Progressive brightening of the core region
in radio was also seen along the VHE flare development. This is an indication that 
the $\gamma$-ray emission originates from a region close to the core rather than from more distant regions.
%Figure~\ref{fig_M87-1} shows the correlated light curves in $\gamma$-rays, X-rays and radio.

%\begin{figure}[!t]
  %\vspace{5mm}
  %\centering
  %\includegraphics[width=2.5in]{icrc0781_fig01.epsi}
  %\caption{Combined M~87 light curves from 2007 to 2008. (A): VHE $\gamma$-ray fluxes
  %($E>0.35$~TeV, nightly average), showing the H.E.S.S., MAGIC and VERITAS data. 
  %(B): Chandra X-ray measurements (2-10~keV) 
  %of the nucleus and the knot HST-1~\cite{Harris2009}. (C): Flux densities from the 43~GHz VLBA 
  %observations. The shaded horizontal area indicates the range of fluxes from the nucleus before the 2008 flare. 
 % Whereas the flux of the outer regions of the jet does not change substantially, 
 % most of the flux increase results from the region around the nucleus. For further details see~\cite{M87Science}}
  %\label{fig_M87-1}
 %\end{figure}

In April 2010, during the seasonal monitoring of M~87, VERITAS detected another flare with peak flux 
of $\sim$20\% of the Crab Nebula flux. During the six-month observation period, M~87 was detected 
at a level of 25.6$\sigma$ above the background, with an average flux above 350~GeV equivalent 
to 5\% of the Crab Nebula flux. Dedicated analysis in 20-minute bins has been performed on the 
April 2010 flaring episode. A spectral analysis has been done on three different phases of the
flaring episode: the rising phase, the peak and the falling phase. A power-law fit has been applied to each phase,
showing a hint of spectral variability: $\Gamma_\mathrm{rise}=2.60\pm0.31$, 
$\Gamma_\mathrm{peak}=2.19\pm0.07$, $\Gamma_\mathrm{fall}=2.62\pm0.18$.

 \begin{figure}[!t]
  \vspace{5mm}
  \centering
  \includegraphics[width=2.8in]{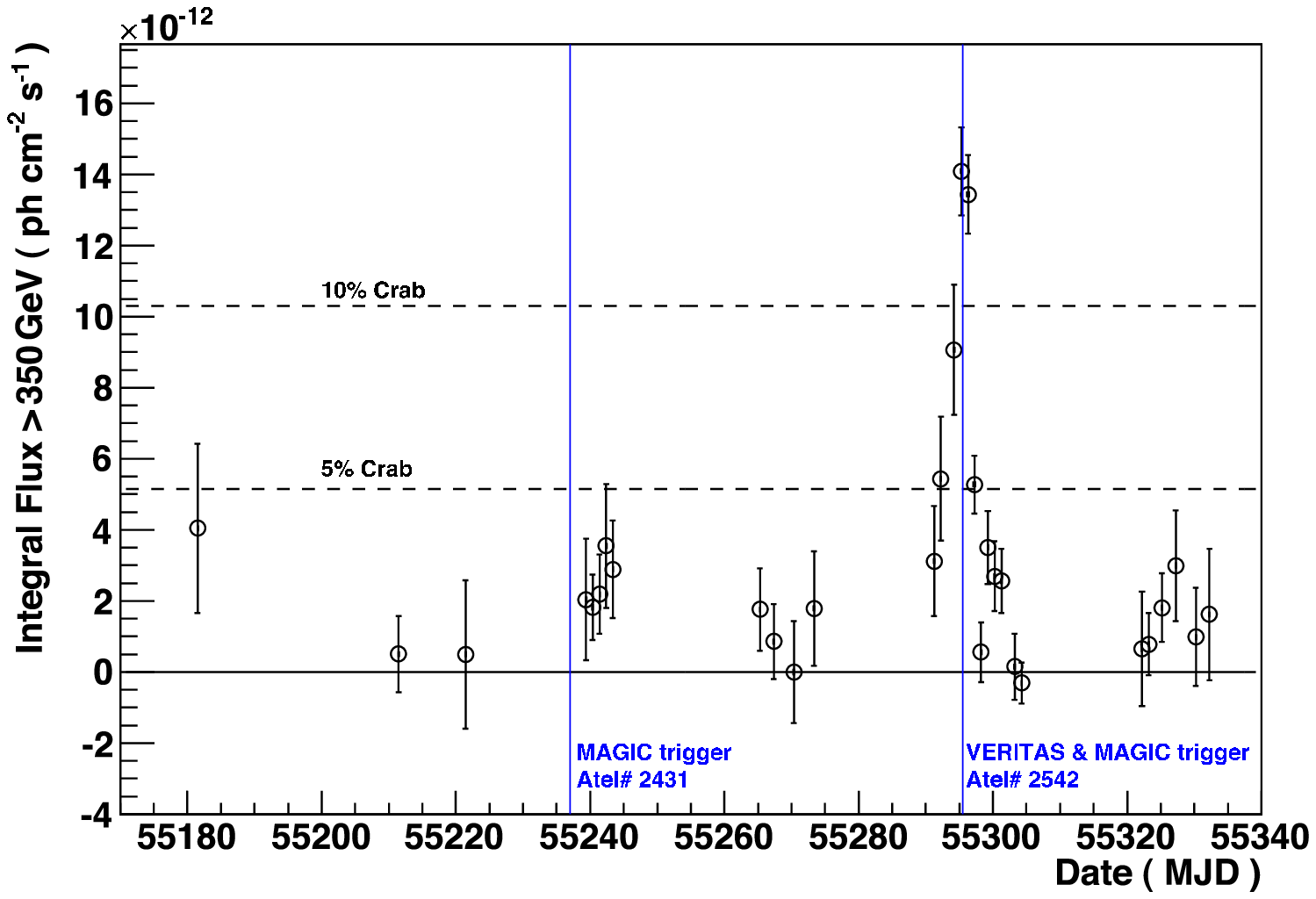}
  \includegraphics[width=2.8in]{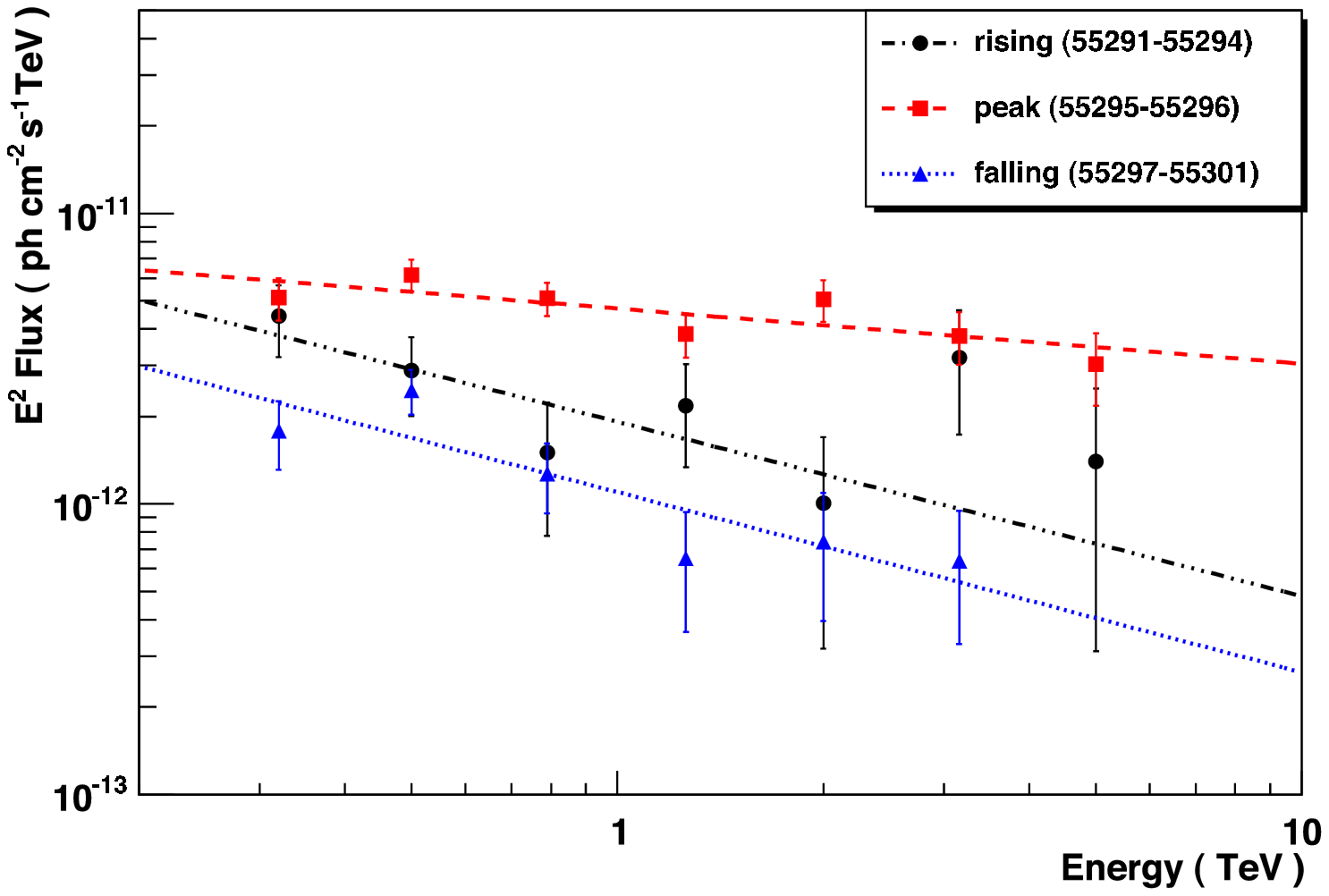}
  \caption{\emph{(upper plot)} VERITAS light curve of the 2010 seasonal monitoring campaign.
  \emph{(lower plot)} Spectral analysis for three phases of the April 2010 flaring event: rising phase ( circles),
  peak (squares) and decreasing phase (triangles).}
  \label{fig_M87-2}
 \end{figure}

Other radio galaxies observed by VERITAS include 3C~111 and NGC~1275. 
A preliminary analysis of 11~hr of quality-selected data of 3C~111 results in a flux upper limit
of $\sim3$\% Crab flux above 200~GeV.
NGC~1275 is an unusual early-type galaxy located in the center
of the Perseus cluster. Its radio emission is core dominated, but emission lines are also seen, making
it difficult to classify it according to the Faranhoff \& Riley (FR) classification~\cite{FR}. In Fall 2008 the
\emph{Fermi} $\gamma$-ray space telescope reported the detection of $\gamma$-ray
emission from a position consistent with the core of NGC~1275. VERITAS observed the core region of 
NGC~1275 for about 11~hr between 2009 January 15 and February 26, resulting in 7.8~hr of quality-selected
live time. No $\gamma$-ray emission is detected above the analysis energy threshold of $\sim$190~GeV,
resulting in a flux upper limit incompatible with the extrapolation of the \emph{Fermi-LAT} spectrum.
Under the assumption of a SSC emission mechanism, the VERITAS result suggests the presence of a cutoff
in the sub-VHE energy range~\cite{Galante}. The detection in Summer 2010 of VHE $\gamma$-ray emission 
by MAGIC~\cite{MAGIC1275} has eventually included NGC~1275 among the few interesting radio galaxies
for future VHE investigation.

\subsection{Clusters of Galaxies}

Observation of clusters of galaxies is done unavoidly during the observation of many radio galaxies. 
This is the case for NGC 1275 and M 87 where the Perseus and Virgo clusters respectively are observed during the 
radio galaxy observation. However, up to now a dedicated study of the clusters themselves
 has been done only on the 
Coma cluster. The Coma cluster is a nearby cluster of galaxies which is well studied at all 
wavelengths~\cite{Neumann2003}. 
It is at a distance of 100~Mpc ($z=0.023$) and has a mass of $2 \times 10^{15}\; M_{\odot}$. 
Its X-ray and radio features suggest the presence of accelerated electrons in the intergalactic medium 
emitting non-thermal radiation. Beside relativistic electrons, there may also be a population of 
highly energetic protons. Both high energy electrons and protons are known to be able to produce VHE photons.
A total of 19 hr of data have been recorded between March and May 2008.
No evidence for point-source emission was observed within the field of view and a preliminary upper limit of 
$\sim$3\% of the Crab flux is given for a moderately extended region centered on the core~\cite{Perkins}.

\subsection{Starburst Galaxies}

M~82 a prototype small starburst galaxy, located approximately 3.7~Mpc  from Earth, 
in the direction of the Ursa Major constellation. M~82 is gravitationally interacting 
with its nearby companion M~81. This interaction has deformed M~82 in such a way that an active starburst 
region in its center with a diameter of $\sim$1000 light years has been developed~\cite{Yun1994,Volk1996}.
Throughout this compact region stars are being formed at a rate approximately 10 times faster 
than in entire ``normal" galaxies like the Milky Way. Hence the supernovae rate is 0.1 to 0.3 per 
year~\cite{Kronberg1985,Fenech2008}.
The high star formation rate in M 82 implies the presence of numerous shock waves in 
supernova remnants and around massive young stars. Similar shock waves are known to 
accelerate electrons to very high energies, and possibly ions too. 
The intense radio-synchrotron emission observed in the central region of M 82 suggests a very 
high cosmic-ray energy density, about two orders of magnitude higher than in the Milky Way~\cite{Rieke1980}.
Acceleration and propagation
of cosmic rays in the starburst core are thus expected to be responsible for VHE $\gamma$-ray
emission.
Theoretical predictions include significant contributions from both leptonic and hadronic particle interactions.
Cosmic-ray ions create VHE gamma rays through collisions with interstellar matter,
producing $\pi^0$ which decay into $\gamma$-rays. Alternatively, accelerated cosmic-ray electrons may
inverse-Compton scatter ambient X-ray photons up to the VHE 
range~\cite{Volk1996,Pohl1994,Persic2008,deCea2009}. 

\begin{figure}[!t]
  \vspace{5mm}
  \centering
  \includegraphics[width=2.8in]{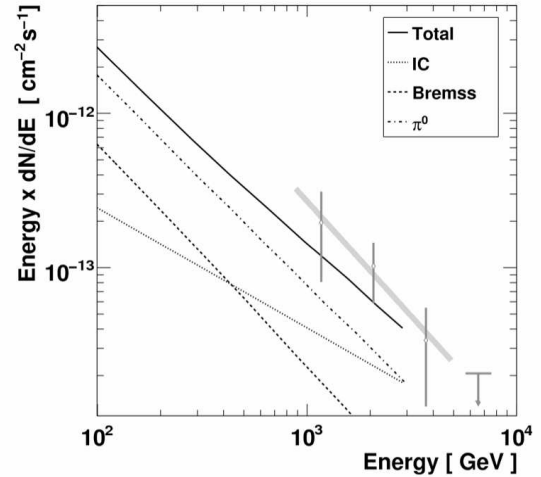}
  \caption{VERITAS $\gamma$-ray flux of M~82 compared to theoretical predictions. 
 The data are given by open diamonds with 1$\sigma$ statistical error bars, and is fitted ($\chi^2/ndf=0.1$) 
 with a power-law function (thick gray line), $\mathrm{d}N/\mathrm{d}E \propto (E / 1\mathrm{TeV})^{-\Gamma}$, 
 with $\Gamma = 2.5 \pm 0.6_\mathrm{stat} \pm 0.2_\mathrm{syst}$. 
 The VERITAS flux upper limit (99\% confidence level~\cite{Rolke2000}) shown at ~6.6 TeV is above the extrapolation of the fitted power-law function at these energies. 
 The thin lines represent a recent model~\cite{Persic2008} for the $\gamma$-ray emission from M~82. 
 The thin solid line is the total emission predicted. The dashed lines represent components from $\pi^0$ decay, 
 and from radiation from cosmic-ray electrons through IC scattering and Bremsstrahlung. 
 The markedly different spectral slopes of these dominant components should be noted.}
  \label{fig_M82}
 \end{figure}

VERITAS observed M~82 for a total of ~137 hours of quality-selected live time between 
January 2008 and April 2009 at a mean zenith angle of 39$^\circ$.
An excess of 91 gamma-ray-like events ($\sim$0.7 photons per hour) are detected for a total
4.8$\sigma$ statistical post-trials significance above 700~GeV. The observed differential 
VHE gamma-ray spectrum is best fitted using a power-law function with a photon index 
$\Gamma = 2.5 \pm 0.6_\mathrm{stat} \pm 0.2_\mathrm{sys}$ (fig~\ref{fig_M82}).
Comparison
of the VERITAS VHE spectrum with predictions of the theoretical models supports a hadronic
scenario as the dominant process responsible for the VHE emission~\cite{Acciari2009}.

\subsection{Globular Clusters}

M~15 (aka NGC~7078) is a very compact ($R_c=0.19$~pc) globular cluster located at a distance of 9.4~kpc.
It belongs to the class of ``core-collapsed" globular clusters and is thought to contain a $\sim2000$~M$_\odot$
black hole in its center. M~15 contains 7 known millisecond pulsars and a low-mass X-ray binary (LMXB) system.
That are expected to emit $\gamma$-rays. Pulsars produce relativistic
strongly magnetized winds which, when colliding with other winds, create relativistic shocks. Models presented
in~\cite{Bednarek2007} consider the case of relativistic leptons injected by the pulsar in such relativistic shocks with a power-law
energy distribution. 

\begin{figure}[!t]
  \vspace{5mm}
  \centering
  \includegraphics[width=2.8in]{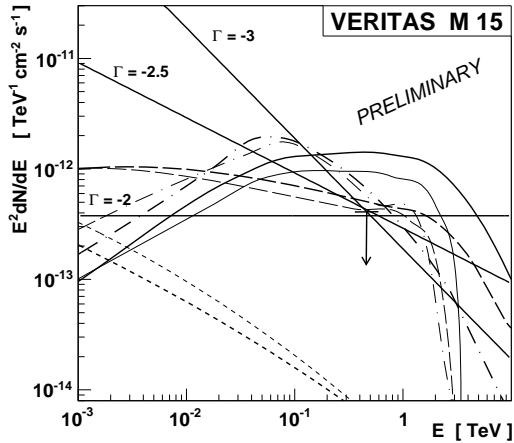}
  \caption{VERITAS 95\% c.l. upper limit on M~15. The three power-laws  represent the
  the three different funtions used to calculate the decorrelation energy. It can be seen that the dependency
  of the flux upper limit on the assumed spectral index is minimal at the decorrelation energy. Models representing
  different leptonic spectra~\cite{Bednarek2007} 
  are also shown: 30~TeV cutoff \emph{(thick lines)}; 3~TeV cutoff \emph{(thin lines)}; 
  $\Gamma=-2.1$, $E_\mathrm{min}=100$~GeV \emph{(solid lines)}; 
  $\Gamma=-2.1$, $E_\mathrm{min}=1$~GeV \emph{(long-dashed lines)}; 
  $\Gamma=-3$, $E_\mathrm{min}=100$~GeV \emph{(long-dashed-dotted lines)}; 
  $\Gamma=-3$, $E_\mathrm{min}=1$~GeV \emph{(short-dashed lines)}.}
  \label{fig_M15}
 \end{figure}

VERITAS observed the globular cluster M~15 between June 10 and June 21,
2010 for a total of 6.2 hours of good-quality live time. The data have been processed with the standard
analysis procedure and cuts optimized for a weak Crab-like source (index $\Gamma=-2.5$, 3\% Crab
Nebula flux). No signal is detected at the cluster's core coordinates. A 95\% confidence level 
upper limit using the Rolke algorithm~\cite{Rolke2000} 
is calculated assuming a Gaussian background. The differential limit
is calculated at the \emph{decorrelation} energy, i.e. the energy that minimizes the dependency of the
flux upper limit on the assumed spectral index. Figure~\ref{fig_M15} shows the VERITAS
upper limit superimposed on the models explained in~\cite{Bednarek2007}. 
Three functions used to calculate the decorrelation
energy are shown as well. It can be seen that the 95\% confidence level VERITAS upper limit places constraints
on the possible scenarios for the spectral energy distribution of the accelerated leptons in the pulsar winds.

\section{Conclusions}

The VERITAS extragalactic non-blazar science program is well established.
In addition to complementing the blazar program on the study of the AGN physics, non-AGN sources are also 
investigated. A dedicated VHE study on the Coma cluster of galaxies resulted in a flux 
upper limit on the extended region centered on the core.
The first detection of $\gamma$-ray emission from a starburst galaxy established a connection
between cosmic-ray acceleration and star formation. 
Observation of globular clusters provides 
constraints on predictions of $\gamma$-ray emission from ms pulsars.\\

This research is supported by grants from the US Department of Energy, the US National Science Foundation, 
and the Smithsonian Institution, by NSERC in Canada, by Science Foundation Ireland, and by STFC in the UK. 
We acknowledge the excellent work of the technical support staff at the FLWO and at the collaborating 
institutions in the construction and operation of the instrument.

%\vspace{\baselineskip}

\clearpage

\end{document}